\documentclass[10pt,conference]{IEEEtran}

\usepackage{graphicx} 
\usepackage{balance}
\usepackage{booktabs} 
\usepackage{longtable}
\usepackage{stfloats} 
\usepackage{tabularx}
\usepackage{url}
\usepackage{url}
\usepackage{amsmath}
\usepackage{listings}
\usepackage{subcaption}
\usepackage{xspace}
\usepackage{float}
\usepackage{array}

\usepackage{siunitx} 
\usepackage[table]{xcolor}
\usepackage[framemethod=tikz]{mdframed}

\newcommand{\toolname}{\textsc{bFinder}\xspace}

\usepackage{multirow}
\definecolor{codegreen}{rgb}{0,0.6,0}
\definecolor{codegray}{rgb}{0.5,0.5,0.5}
\definecolor{codepurple}{rgb}{0.58,0,0.82}
\definecolor{backcolour}{rgb}{0.95,0.95,0.92}

\lstdefinestyle{javastyle}{
    backgroundcolor=\color{backcolour},   
    commentstyle=\color{codegreen},
    keywordstyle=\color{magenta},
    numberstyle=\tiny\color{codegray},
    stringstyle=\color{codepurple},
    basicstyle=\ttfamily\footnotesize,
    breakatwhitespace=false,         
    breaklines=true,                 
    captionpos=b,                    
    keepspaces=true,                 
    numbers=left,                    
    numbersep=5pt,                  
    showspaces=false,                
    showstringspaces=false,
    showtabs=false,                  
    tabsize=2
}

\usepackage{tcolorbox}
\tcbuselibrary{most}
\newtcolorbox{rqbox}[2][]{%
    colback=black!5,
    colframe=black!40,
    colbacktitle=black!20,
    coltitle=black,
    fonttitle=\bfseries, 
    title=#2,
    enhanced,
    attach boxed title to top left={yshift=-2mm, xshift=-1mm}, %
    #1%
}

\title{Identifying Behavioural Gaps in Unit Tests Using Documentation}

\title{Beyond Coverage and Kill Scores:\\ An Empirical Study of Behavioural Gaps}

\title{Beyond Coverage and Kill Scores: Empirically\\ Measuring Behavioural Gaps in Test Suites}

\title{Beyond Coverage and Kill Scores: Empirically\\ Measuring Test Suite Behavioural Gaps}

\author{
    \IEEEauthorblockN{Partha Protim Paul and Reid Holmes}
    \IEEEauthorblockA{University of British Columbia, Vancouver, Canada\\
    \{pppaul, rtholmes\}@cs.ubc.ca}
}

\begin{document}
\maketitle

\begin{abstract}


Traditional test adequacy metrics measure a system's implementation, not whether it adheres to its expected behaviour. 
While developers rely heavily on code coverage and mutation testing to assess test suite quality, these metrics are fundamentally implementation-centric and cannot detect gaps between
what the code is \textit{expected} to do and what it \textit{actually} does.
Unfortunately, there has been no way to reliably detect these discrepancies; in this paper we introduce an automated proof-of-concept approach to investigate these gaps. 
The approach extracts expected method-level behaviours from natural language documentation and source code, maps them to existing test cases, and identifies gaps between \textit{expected} and \textit{validated} behaviours. 
We evaluate the approach across ten popular open-source Java libraries comprising 8,922 methods, extracting 20,729 behaviours with 93.1\% precision. 
Our empirical analysis conservatively estimates that 17.5\% of detected expected behaviours remain entirely untested, which we term as the test suite's \textit{behavioural gap}. 
To determine if these gaps are merely an artifact of human-driven testing, we evaluate state-of-the-art automated test generators 
(\textsc{EvoSuite} / \textsc{ASTER}), finding that they similarly fail to validate at least 20.6\% / 27.1\% 
of detected expected behaviours. We further demonstrate that behavioural gaps are not predicted by traditional structural metrics: the majority of untested behaviours occur in methods that already have high line coverage, and over half persist in methods with high mutation kill score. 
These results suggest behavioural coverage acts as an independent dimension of test 
suite adequacy that can complement traditional structural metrics.
\end{abstract}


\section{Introduction}
\label{sec:intro}

Early testing approaches directed validation effort on expected program behaviour~\cite{Howden1978FunctionalPT, richardson89}, in contrast to the current more common approaches that examine implementation structure~\cite{myers1979art}. 
In practice, expected behaviours are often expressed through informal documentation, edge cases, and API contracts, making behavioural completeness difficult to evaluate systematically~\cite{hossain2025doc2oracllinvestigatingimpactdocumentation, 10.1109/ICST.2012.110}. 
As a result, software testing research and practice have largely converged on implementation-centric structural proxies for test quality~\cite{howden1976reliability, Budd1978Design, zhu1997software}, with
code coverage and mutation kill scores 
commonly being used to evaluate 
how complete a test suite is~\cite{papadakis2019mutation, 10.1145/2568225.2568278, 10.1145/2635868.2635929}.

Unfortunately, structural test measures have introduced shortcomings that decouple test metrics from true behavioural intent. 
As these metrics validate correctness relative to the existing implementation, they are inherently blind to faults of omission: missing logic or unhandled requirements can still yield high structural scores~\cite{juristo2003functional, lawrance2005how}. 
These metrics also focus on code execution rather than behavioural validation; a test suite can maximize coverage or kill scores while lacking the robust oracles needed to validate expected behaviours, not just directly measuring the code as it was implemented~\cite{zhang2015assertions, petrovic2021does}. 
For instance, as we show in Section~\ref{sec:example}, a test achieving 100\% coverage can still leave an entire documented behaviour untested, leaving any regression against it undetected. 
Early software engineering theory warned that structural execution cannot guarantee that a program fulfills its expected behaviours~\cite{miller1963systematic, goodenough1975toward, hamlet1994foundations}; modern empirical evaluations confirm that actively optimizing for these code-centric thresholds yields highly optimized test suites with degraded fault-detection capabilities~\cite{10.1145/2568225.2568271, gay2015risks}.


The primary advantage of structural measures is that they are easy to collect as they map directly to the source code, which is amenable to automated analyses. 
This contrasts with expected behaviours, which are 
often encoded within natural language documentation, providing an abstract, implementation-independent description of what a method should do~\cite{hossain2025doc2oracllinvestigatingimpactdocumentation}, and are not currently directly measurable like coverage and kill scores. 
While measuring whether a test suite validates documented behaviours has remained challenging because it requires linking natural language documentation to the behaviours evaluated by test cases~\cite{lee2025metamon}, 
recent advances in large language models (LLMs) now make this form of analysis feasible~\cite{Endres2023CanLL}. 
This paper therefore examines: \textit{To what extent can documentation-derived expected behaviours augment structural metrics in revealing  
inadequacies 
in existing test suites?}
We investigate this question empirically through four research questions around measuring behavioural coverage in software testing.


\textbf{RQ1 (Feasibility): With what precision can expected behaviours be extracted from source code and documentation?}
Before we can evaluate test adequacy, we must first determine if expected method-level behaviours can be extracted from natural language and code with sufficient precision for large-scale analysis. To this end, we introduce \toolname{}, a proof-of-concept that extracts expected behaviours, maps them to corresponding test cases, and identifies untested behaviours. Across ten popular open-source projects containing 8,922 documented methods, \toolname{} surfaces 20,729 distinct behaviours with a validated precision of 93.1\%.


\textbf{RQ2 (Prevalence): To what extent do mature, developer-written test suites exhibit behavioural gaps?}
To understand the prevalence of behavioural gaps in real-world systems, we analyze the test suites of the same ten projects. We define a \textit{behavioural gap} as a documented behaviour that is not validated by any existing test case. Our analysis conservatively finds that at least 17.5\% of detected expected behaviours are not covered by these existing, well-maintained test suites. Furthermore, a qualitative inspection of these gaps indicates that these behaviours are rarely inherently untestable, suggesting a systematic shortfall in current validation practices.


\textbf{RQ3 (Generalization): Do behavioural gaps persist when test suites are automatically generated?}
To determine whether behavioural gaps are inherent to manual test authoring or a systemic limitation of structure-guided testing practices, we evaluate whether they persist in automatically generated test suites. We apply two state-of-the-art test generation approaches, \textsc{EvoSuite}~\cite{10.1145/2025113.2025179} and \textsc{ASTER}~\cite{Pan2024ASTERNA}, to four of our subject systems. We find that 27.1\% of behaviours are missed by \textsc{ASTER} and 20.6\% by \textsc{EvoSuite}. These results suggest that behavioural gaps are not just an artifact of human-written code, but a fundamental challenge for both human and automated testing strategies.


\textbf{RQ4 (Implication): Do behavioural gaps alias traditional structural metrics?}
We hypothesize that behavioural gaps are not simply proxies for low coverage or mutation scores, but represent a distinct dimension of test adequacy. Across the untested behaviours identified in our developer-written suites, we find that the majority of untested behaviours occur in methods that already achieve high structural scores. Most fall in methods exceeding 90\% line coverage, and over half in methods exceeding 90\% mutation kill scores. 
These findings indicate that behavioural gaps can identify validation weaknesses in code that is `comprehensively' tested, suggesting that behavioural gaps can complement analysis of structurally-strong suites.


This paper makes the following key contributions:
\begin{itemize}
    \item \textbf{Empirical Analysis of Behavioural Gaps:} We conduct an empirical study across ten popular open-source Java libraries, demonstrating that significant behavioural gaps exist between the documented and validated behaviours of real-world methods, even in suites with high structural coverage.
    
    \item \textbf{Generalization to Automated Testing:} We present a direct comparison against two state-of-the-art automated test generation approaches, confirming that these behavioural gaps persist even in fully-generated test suites.

    \item \textbf{Behavioural Coverage as a Distinct Dimension:} We show that behavioural gaps do not alias line coverage or mutation kill score. Gaps persist in 38.2\% of methods with perfect line coverage and 29.6\% with a perfect mutation kill score, establishing behavioural coverage as a complementary dimension of test adequacy. 
    
    \item \textbf{Tool \& Artifacts:} We provide a publicly available replication package, including \toolname{} and all identified behavioural gaps, to enable researchers to reproduce our results and extend this analysis to other software systems.\footnote{\toolname{}: \url{https://anonymous.4open.science/r/behavioural-coverage}}
\end{itemize}

\noindent Section~\ref{sec:example} provides a motivating example showing how behaviours are missed, despite being covered. Section~\ref{sec:related} describes related work. \toolname{} is introduced in Section~\ref{sec:approach}. Section~\ref{sec:eval} describes empirical study and its findings.
Section~\ref{sec:summary} concludes.

\section{Motivating Example}
\label{sec:example}

The \texttt{Triple.emptyArray()} method from the \texttt{commons-lang} library is shown at the top of Figure~\ref{triple:emptyArray}. 
From the documentation, a developer would expect the method to provide two main behaviours: First, the description and the method name itself make it clear that the method should return an empty array. Second, the \texttt{@return} block reinforces that the method should always return the same array instance, since it is defined as a singleton.

\begin{figure}[h]
\begin{lstlisting}[language=Java]
/** Returns the empty array singleton that can   
 * be assigned without compiler warning.
 * 
 * @param <L> the left element type.
 * @param <M> the middle element type.
 * @param <R> the right element type.
 * @return the empty array singleton that can
 *  be assigned without compiler warning.
 * @since 3.10
 */
public static Triple<L, M, R>[] emptyArray() {
  return (Triple<L, M, R>[]) EMPTY_ARRAY;
}

// Only developer-written test for emptyArray
@Test
void testEmptyArrayGenerics() {
  Triple<Integer, String, Boolean>[] empty =
    Triple.emptyArray();
  assertEquals(0, empty.length);
}
\end{lstlisting}
\caption{Documentation, implementation, and developer-written test for \texttt{emptyArray()} from the \texttt{commons-lang} project.}
\label{triple:emptyArray}
\end{figure}

The only developer-written test case for \texttt{emptyArray} is shown at the bottom of Figure~\ref{triple:emptyArray}. While this test case achieves 100\% coverage for the \texttt{emptyArray} implementation, it only validates the first expected behaviour: that the returned array is actually empty.
The existing test suite exhibits a \emph{gap} between the expected behaviours and existing suite: the suite fails to validate that the returned array is a singleton. 
As a result, if a regression were introduced that changed the method to return a new empty array each time, this change would go undetected and could introduce new faults to users of \texttt{emptyArray()} who relied on the method returning a singleton instance of the empty array.



This paper seeks to address the gap between the source code and test case above.
While the developer-written test suite appears to be relatively complete from a coverage point-of-view, the test cases do not actually validate all of the behaviours a developer using this method might expect they could rely upon.
Our approach, called \toolname, takes as input the source code and documentation for a method, along with its existing test suite, and identifies these behavioural gaps between existing tests and expected behaviours.
These gaps are actionable: developers can resolve them by strengthening their test suite to directly address the specifically-identified weakness in their existing tests.
This gap can be easily addressed by a single \texttt{assertSame(..)} assertion across two invocations to validate that the returned array is the same instance for multiple calls.

The remainder of this paper describes an approach for detecting behavioural gaps, evaluates the accuracy of this technique, and describes the kinds of behaviours that are frequently missed by existing developer-written test suites.


\section{Related Work}
\label{sec:rw}
\label{sec:related}

Prior work has studied test suite completeness through structural metrics and code-centric gap analysis, and has used documentation to guide test generation. However, whether a semantic gap exists between the behaviours the developer documented and the behaviours their test suites actually validate has not been studied. In the following subsections, we discuss how each of these areas provides context to our study.

\subsection{Structural Test Measures}

Code coverage and mutation kill score are the two most widely adopted proxies for test suite quality in both research and practice~\cite{papadakis2019mutation, 10.1145/2568225.2568278, 10.1145/2635868.2635929}. Coverage measures whether source lines are executed by tests, while mutation kill score assesses whether those executions are backed by meaningful assertions~\cite{10.1145/1062455.1062530}. Mutation kill score has been shown to correlate significantly with real fault detection independently of coverage, making it the best available proxy for test suite effectiveness~\cite{10.1145/2635868.2635929}.
Together, they are often treated as complementary signals of test suite completeness. However, other studies have shown that coverage correlates only weakly with fault detection once test suite size is controlled for~\cite{10.1145/2568225.2568271}, and mutation testing incurs significant computational cost that limits its scalability in practice. 
Fundamentally, both metrics are implementation-centric, measuring properties of the code as written rather than whether the code fulfills its documented behaviours. 

\subsection{Complementing Structural Testing} 

Several prior projects have studied the notion of \emph{testing gap} from a code-centric perspective. They have mainly focused on situations where tests execute code but fail to meaningfully check its behaviour. Schuler \textit{et al.}'s checked-coverage technique identifies code whose effects are not validated by the assertions in the tests~\cite{schuler2013checked}. Similarly, mutation-testing research has explored pseudo-tested methods, where mutants can be introduced without causing test failures, indicating weak or ineffective tests~\cite{vera2019comprehensive}. More recently, Maton \textit{et al.} introduced \textsc{GapGrep}, which connects checked coverage and mutation testing to identify execution gaps in test suites~\cite{maton2025tests}. 

These approaches provide insight into whether the executed code is meaningfully asserted. However, they remain fundamentally code-centric and do not consider whether tests validate the behaviours developers explicitly describe in documentation. Fraser \textit{et al.} 
recognized this limitation early, proposing behavioural adequacy as a criterion grounded in whether test executions can reproduce an accurate model of system behaviour~\cite{10.1109/ICST.2012.110}. 
Despite this early recognition, to the best of our knowledge, no work has characterized how large this gap is between documented intent and what existing developer-written test suites actually validate. We address this gap empirically, using LLMs as a measurement instrument rather than a generation engine, to quantify the lower bound on how much of the documented intent of real-world Java libraries goes unvalidated in practice.




\subsection{Automated Test Generation}

Automated test generation tools target structural coverage through various strategies: search-based approaches (e.g., EvoSuite~\cite{10.1145/2025113.2025179}), symbolic execution~\cite{kurian2023automatically}, constraint solving~\cite{blasi2022call}, fuzzing~\cite{10.1145/3611643.3616327, pacheco2007feedback}, and LLM-based methods such as ChatUniTest~\cite{chen2024chatunitest} and ASTER~\cite{Pan2024ASTERNA}.

\begin{figure*}[ht!]
    \centering
    \includegraphics[width=0.8\textwidth,keepaspectratio]{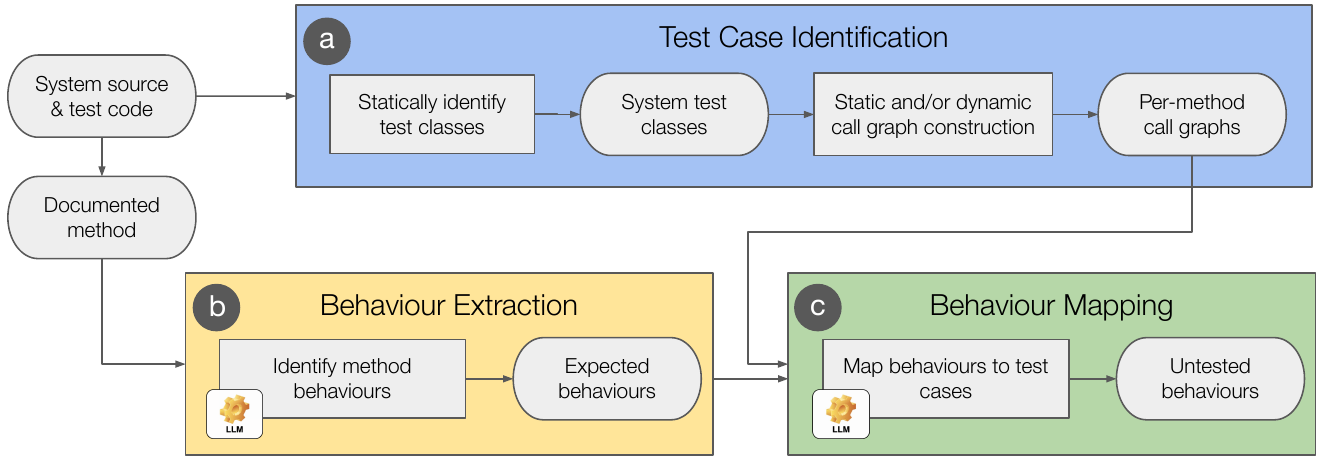}
    \caption{Process \toolname follows to identify any untested behaviours for a given documented method and the system's existing developer-written test suite.\vspace{-1em}}
    \label{fig:tool}
\end{figure*}

Some tools use method documentation to guide test generation. 
Toradocu extracts \textit{exception}-related conditions from structured Javadoc tags to generate executable oracles~\cite{10.1145/2931037.2931061}. 
JDoctor extends this approach to infer pre- and post-conditions from \textit{@param}, \textit{@return}, and \textit{@throws} tags~\cite{10.1145/3213846.3213872}, while Judot uses these inferred contracts to guide search-based test generation~\cite{denaro2025automatedtestgenerationprogram}. 
However, JDoctor's reliance on structured tags limits its reach to methods with 
formal documentation, leaving behaviours expressed in narrative text outside its scope. DISTINCT uses LLMs to generate tests by using natural language descriptions and branch-level coverage analysis~\cite{zhang2025frameworkcreatingnonregressivetest}. 

More recently, LLMs have been applied to tasks that bridge natural language and code. Endres~\textit{et al.}~\cite{Endres2023CanLL} showed that Javadoc can be translated into executable postconditions, and tools like DocPrism~\cite{xu2025docprismlocalcategorizationexternal} use LLMs to detect code-documentation inconsistencies. 
DocPrism showed that although documentation is sometimes viewed as unreliable, documentation errors are relatively infrequent ($\sim$10\%) and are typically detectable~\cite{xu2025docprismlocalcategorizationexternal}, suggesting documentation can be relied upon in practice.

Despite this progress, these tools treat documentation as a blueprint for creating new tests or oracles from scratch, but do not evaluate whether the existing developer-written test suite already covers the documented behaviours. As a result, even with documentation-aware generation, the extent to which existing test suites fail to validate their own documented intent remains an open empirical question. 
In this paper we empirically measure behavioural gaps in both developer-written test suites and in suites created by automated test generation approaches.

\section{\toolname{}: Detecting Untested Behaviours}
\label{sec:approach}

We built the \toolname{} proof-of-concept to investigate the prevalence of behavioural gaps in real-world Java systems.
The high-level process \toolname{} follows is shown in Figure~\ref{fig:tool}.
First, the approach creates per-method call graphs to link source methods to the test cases that validate them.
Second, the behaviour extractor uses an LLM-based approach to identify the expected behaviours for any documented method.
Finally, the method's expected behaviours are mapped to the behaviours validated by the method's test cases to determine which behaviours remain untested by the existing test suite. 
\toolname{} works on Java (v8+) systems using the JUnit test library (v4+).
\toolname{} is available in the replication package.



\subsection{Test Case Identification (Figure~\ref{fig:tool}(a))}
\label{sec:approach:tests}

To determine which behaviours of a method are validated by the test suite, \toolname{} first identifies the set of test cases that invoke each method under test. 
The approach statically scans the repository for JUnit \texttt{@Test} and \texttt{@ParameterizedTest} annotations to enumerate all test classes, then constructs a full-system call graph capturing how each test case interacts with the methods under test.

By default, \toolname{} uses dynamic analysis: each test class is executed in isolation using an AspectJ\footnote{AspectJ: \url{https://github.com/eclipse-aspectj/aspectj}}-based monitoring agent that records every method entry and exit. 
Dynamic analysis is preferred because it captures exact runtime execution paths, avoiding the over- and under-approximation inherent in static analysis. 
The cost of this dynamic analysis is approximately 2x the test execution time compared to an uninstrumented test suite.

Because the per-class call graph contains detail irrelevant to any individual method under test, \toolname{} prunes the graph into a dedicated call graph per method. 
A breadth-first traversal finds all paths that eventually invoke the target method; the surviving nodes are then statically enriched with each method's implementation, Javadoc, and referenced class-level variables. 
The result is one enriched entry method under test, which serves as the primary input to the behaviour mapping step (Section~\ref{sec:approach:mapping}).

\toolname{} also supports several specializations for flexible use and increased performance.
Analysis scope can be narrowed from the full system to a single directory, class, or method, substantially reducing runtime. 
For systems with straightforward test suites, static call-graph tracing can replace dynamic instrumentation entirely resulting in negligible test case identification overhead.
Finally, \toolname{} integrates naturally into CI/CD pipelines: given a diff, it can update call-graph knowledge for only the affected source and test methods rather than recomputing all test-to-code links from scratch.

\subsection{Extracting Behaviours (Figure~\ref{fig:tool}(b))} 
\label{sec:approach:behaviours}

\toolname{} generates a list of expected behaviours for every source method in the system that is accompanied by a method-level Javadoc comment.
Our proof-of-concept focuses on documented methods because developers use documentation as a flexible way to surface a method's expected behaviours without inspecting its implementation~\cite{hossain2025doc2oracllinvestigatingimpactdocumentation}.
For each such method, \toolname invokes an LLM with some guidance, the method signature, implementation, and Javadoc. 
%
The output of this is a list of expected behaviours (denoted $EB_1, EB_2,\dotsc$). Each expected behaviour is typically expressed as a single sentence and describes a property or guarantee a developer would reasonably expect from the method based on its documentation. 
These behaviours form the reference set against which test cases that validate the method are evaluated. 
For the code given in Figure~\ref{triple:emptyArray}, \toolname{} returns two expected behaviours: ``Returns an empty array with no elements``, and ``Returns the 
same singleton array instance every time.'' 


LLMs are well-suited to behaviour extraction because they can interpret unstructured natural language and summarize it into discrete, testable behaviours~\cite{Wong_2023}.
This has been leveraged by prior approaches; for example,
JDoctor~\cite{10.1145/3213846.3213872} and Toradocu~\cite{10.1145/2931037.2931061} extract formal contracts only from structured Javadoc tags (\texttt{@param}, \texttt{@return}, \texttt{@throws}). 
In this work, we focus on the entire Javadoc block, including any narrative text the developer thought would be important for understanding a method.
Our goal in this phase is to surface the set of behaviours a developer might reasonably expect based on the full documentation of a method, rather than to infer formal contracts. 

\begin{table*}[b]
\centering
\caption{Dataset used from \texttt{Defects4J} along with their key properties and details of their developer-written test suites.} 
\label{tab:repositories}
\begin{tabular}{lrrrrr}
        \toprule
        \textbf{Project (Version)} & \textbf{\# Source Methods} & \textbf{\# Test Methods} & \textbf{SLOC} & \textbf{Line Cov.} & \textbf{Mutation Kill Score} \\
        \midrule
        commons-cli (1.10.0-SNAPSHOT) & 214 & 461 & 1965 & 98.2\% & 93\% \\
        commons-codec (1.21.1-SNAPSHOT) & 523 & 1,196 & 4972 & 94.9\% & 92\% \\
        commons-compress (1.29.0-SNAPSHOT) & 997 & 2,254 & 25308 & 88.0\% & 82\% \\
        commons-jxpath (1.4.1-SNAPSHOT) & 98 & 367 & 9,228 & 77.7\% & 75\% \\
        commons-collections (4.5.1-SNAPSHOT) & 1,352 & 1,901 & 14751 & 85.9\% & 87\% \\
        commons-lang (3.18.0-SNAPSHOT) & 2,703 & 4,483 & 16,133 & 96.5\% & 88\% \\
        commons-csv (1.14.2-SNAPSHOT) & 175 & 511 & 1,225 & 99.6\% & 95\% \\
        gson (2.13.3-SNAPSHOT) & 244 & 1,395 & 5,253 & 92.2\% & 91\% \\
        jfreechart (2.0.0-SNAPSHOT) & 2,068 & 2,299 & 51,346 & 56.5\% & 59\% \\
        jsoup (1.21.1) & 548 & 1,442 & 9,778 & 89.6\% & 79\% \\
        \bottomrule
    \end{tabular}
    \vspace{-1.0em}
\end{table*}

\subsection{Behaviour Mapping (Figure~\ref{fig:tool}(c))} 
\label{sec:approach:mapping}

Once the method's expected behaviours and relevant test cases are known, \toolname{} determines which of the expected behaviours are not validated by the existing test cases. 
This phase establishes a traceability link between a method's natural language documentation and concrete test cases that validate its behaviour.

For each relevant test case, \toolname{} uses an LLM to determine which expected behaviour(s) the test case validates.
The model is provided with a set of goals, the method under test and its documentation, the test case method being evaluated, the list of expected behaviours ($EB = \{EB_1, EB_2, \dots, EB_n\}$), and the relevant nodes in the call graph between the method under test and the test case. 
Including the call graph allows the model to consider indirect interactions between the test case and the method under test.
This step produces a set of tested behaviours, denoted $TB = \{TB_1, TB_2,\cdots, TB_n\}$ (by construction, $TB \subseteq EB$), where each element corresponds to an expected behaviour that is validated by the test case.

Finally, \toolname{} emits the set of untested behaviours as $UTB = EB - TB$.
These untested behaviours (UTB) represent gaps between the behaviours derived from the method's documentation and the behaviours validated by the existing test suite. 
The average processing time was $\approx$3--4 minutes per method, including both behaviour extraction and test mapping. On local inference, the primary bottleneck was LLM response latency, accounting for nearly 2--3 minutes for a method with 7--8 associated tests. Token usage averaged 6K--8K per method under test, scaling with the number of associated test cases. 
This relatively small token burden can be processed by commercial LLM services in $<10$ seconds per method.



\section{Evaluation}
\label{sec:eval}


Unit test suites strive to comprehensively validate that the code under test behaves as expected. 
Structural metrics such as line coverage and mutation kill score are widely used to assess this, but they do not capture whether all expected behaviours are actually validated. 
In this section, we investigate whether gaps exist between the behaviours developers document and the behaviours their test suites validate.
%

Specifically, our evaluation seeks to answer the following research questions:

\begin{itemize}
    \item \textbf{RQ1 (Feasibility): With what precision can expected behaviours be extracted from source code and documentation?} 
     Before studying behavioural gaps, we must establish whether automatically-extracted behaviours are meaningful and grounded in the documented intent of each method. This RQ evaluates the quality of behaviour extraction as a precondition for the remaining research questions.
    
    \item \textbf{RQ2 (Prevalence): Do developer-written test suites exhibit behavioural gaps in practice?} 
     This RQ investigates whether behavioural gaps exist in practice, how large they are, and what kinds of behaviours developers commonly overlook in their test suites.
    
    \item \textbf{RQ3 (Generalization): To what extent are auto\-matically-generated test suites prone to behavioural gaps?} 
     This RQ determines whether automatically-generated suites also exhibit behavioural gaps to determine whether these gaps are inherent to manual test writing practice.
    
    \item \textbf{RQ4 (Implication): Do behavioural gaps alias traditional structural metrics?} 
     This RQ examines whether behavioural gaps are captured by existing structural metrics such as line coverage and mutation kill score, or whether they represent a complementary dimension of test suite adequacy.
\end{itemize}

Experimental design decisions all have benefits and drawbacks. Following best-practice advice~\cite{design-tradeoffs}, we report the experimental design decisions and their implications along with our experimental methodology rather than a separate threats to validity section. 
The full replication package, including all identified untested behaviours is publicly available.\footnote{\toolname{}: \url{https://anonymous.4open.science/r/behavioural-coverage}}

\smallskip
\noindent\textbf{LLM Configuration.} All experiments were conducted using the Qwen3-14B model~\cite{qwen3technicalreport}, executed locally on a single NVIDIA A100 80GB GPU with recommended settings (Temperature=0.6, TopP=0.95, TopK=20, and MinP=0).\footnote{\url{https://huggingface.co/Qwen/Qwen3-14B\#best-practices}}

\smallskip
\noindent\textbf{Project Selection}
Evaluated projects were selected from 
Defects4J~\cite{10.1145/2610384.2628055}, a widely used benchmark in software testing research. 
We applied a systematic filtering process to ensure technical consistency and reproducibility. 
We retained only projects that: (1) contain a \texttt{pom.xml} file in the root directory, ensuring Maven-based builds; (2) are not part of the libraries used in our \toolname{} infrastructure; (3) are hosted as monorepos, meaning each repository contains only the library's code rather than a collection of different libraries; and (4) use JUnit version 4 or higher. Applying these criteria yielded ten projects. 
We analyzed all public and protected documented methods that were executed by at least one test case.
A summary of the selected projects is shown in Table~\ref{tab:repositories}.


Coverage was collected with JaCoCo~\cite{jacoco2026}; mutation kill scores were collected using PIT~\cite{pitest2026}. 
Most projects have high coverage and kill scores. Eight of ten projects exceed 80\% line coverage; seven projects exceed 80\% mutation kill score.

\smallskip
\noindent\textbf{Implication:} 
These projects enable studying whether behavioural gaps exist in popular, long-lived projects that have invested significant testing effort.
By investigating behavioural gaps in comprehensively-tested projects, we hope our findings represent a lower bound for missed behaviour.
The project selection methodology limits our analysis to one broad class of project (libraries) and one 
programming language (Java). 
Evaluating the approach on other languages and domains remains for future work.

\subsection{RQ1 (Feasibility): With what precision can expected behaviours be extracted from source code and documentation?}
\label{sec:rq1}


Before quantifying behavioural gaps, we must ensure that the behaviours the \toolname{} proof-of-concept surfaces are valid. 
To this end, we first investigate the precision of the \toolname{} behaviour extraction (Figure~\ref{fig:tool}(b)). 
\toolname{} identified 20,729 behaviours across all documented methods in our dataset.
To evaluate whether these behaviours are correct, 
we adopt a two-step approach: we first establish human consensus labels to determine how consistently behaviours can be labelled from documentation and source code.
We next build an LLM-based judge, compare its labelling performance against the prior human labels, and apply this judge against all identified behaviours to assess the overall precision of \toolname{}.



\smallskip
\noindent\textbf{Behaviour assessment judge:} The judge is an agentic system built on the Ministral 3:14B reasoning model with its default configuration. We use a different model family (Ministral) as the judge from the one used for behaviour extraction (Qwen) to mitigate self-enhancement bias~\cite{zheng2023judging}.
We choose this model for its strong reasoning performance relative to similar-sized models, which is important for behaviour validation~\cite{liu2026ministral}. 
Substituting the model with a more advanced reasoning model should only increase the judge's precision. 

The judge is provided a method behaviour, the method's documentation and implementation, and can query other method implementations as needed via tool use. 
The judge then assesses whether the provided behaviour would be expected. 
Across all projects, the judge labelled 19,829 (95.7\%) behaviours as valid and 900 (4.3\%) as invalid.
Table~\ref{tab:method_behaviour_counts} reports the judge's output alongside \toolname{}'s extracted counts. 

\smallskip
\noindent\textbf{Validating behaviour judge against human labels:}
Recent work has shown that LLMs can effectively label software engineering tasks when properly validated~\cite{ahmed2025can, 10.1145/3797276, zhou2025llm}. We follow this practice and validate our judge through manual labelling before trusting it at scale.
As the judge’s predictions are imbalanced (19,829 valid vs. 900 invalid), a proportional random sample would 
over-sample valid judgments. 
We therefore used a disproportional stratified random sample~\cite{rs11020185} of 100 behaviours, taking 50 from each stratum (valid and invalid) to ensure the minority class is sufficiently represented. 
Within each stratum, we further distribute samples evenly across the 10 projects. 


Two authors, each with several years of Java experience, independently labelled each of the 100 behaviours as \textit{valid} or \textit{invalid} based on the method and its documentation. 
The two labellers reached \emph{substantial} agreement (Cohen's $\kappa=0.77$, Krippendorff's $\alpha=0.79$)~\cite{mchugh2012interrater}.
A post-hoc power analysis ($\alpha=0.05$, power=0.90) indicates that 43 samples are sufficient to detect this level of agreement~\cite{doi:10.1177/02537176261422290}, well within our sample of 100. 
The labellers resolved any disagreements collaboratively to produce a consensus label for each behaviour to serve as a ground truth against which to evaluate the judge.

Comparing the judge's original prediction to these consensus labels yields $\kappa = 0.68$ and $\alpha = 0.68$, again representing \emph{substantial} agreement.
The post-hoc power analysis ($\alpha=0.05$, power=0.90) indicates that a minimum of 89 samples are required to detect this level of agreement.

\begin{table}[h]
\centering
\caption{\toolname{} extracted behaviours and judged validity.}
\label{tab:method_behaviour_counts}
\begin{tabular}{lrrrr}
\toprule
\textbf{Project} & \textbf{\# Behaviours} & \textbf{\# Valid} & \textbf{\# Invalid (\%)} \\
\midrule
commons-cli         & 424 & 402 & 22 (5\%)\\
commons-codec       & 1,218 & 1,158 & 60 (5\%) \\
commons-compress    & 1,707 & 1,614 & 93 (5\%) \\
commons-jxpath      & 183 & 177 & 6 (3\%) \\
commons-collections & 2,816 & 2,681 & 135 (5\%) \\
commons-lang        & 7,719 & 7,380 & 339 (4\%)\\
commons-csv         & 361 & 348 & 13 (3\%) \\
gson                & 554 & 536 & 18 (3\%) \\
jfreechart          & 4,532 & 4,366 & 166 (4\%) \\
jsoup               & 1,215 & 1,167 & 48 (4\%) \\
\midrule
Total               & 20,729 & 19,829 & 900 (4\%) \\
\bottomrule
\end{tabular}
\end{table}

\smallskip
\noindent\textbf{Estimating precision over all extracted behaviours:}
Since the judge serves as our measurement instrument over all behaviours ($AB=20,729$), its precision directly determines our confidence in \toolname{}'s extracted behaviours.
Recall cannot be calculated because expected behaviours lack a discrete ground truth; their decomposition into individual behaviours is inherently subjective. 
We therefore evaluate precision, as determining whether an extracted behaviour is valid constitutes a well-defined binary judgment.
A high precision implies that when \toolname identifies a behaviour, we can be confident it is valid and expected given a method and its documentation. 

We use the consensus labels from the 100 samples to estimate per-stratum validity rates.
Among the 50 samples drawn from the judge's valid stratum, 48 are confirmed valid compared to the human labels ($\hat{p}_{valid} = 0.96$). 
Among the 50 samples from the invalid stratum, 14 are confirmed valid ($\hat{p}_{invalid} = 0.28$). This reflects the judge's conservative tendency to flag some valid behaviours as invalid. 

Specifically, the judge tended to reject behaviours that were implied rather than explicitly stated in the documentation.
For example, flagging a behaviour about negative inputs as invalid when the Javadoc only discussed positive ones. 
The judge also struggled with ambiguous descriptions that human annotators could resolve from context. 
The former reflects strict adherence to explicit documentation, the latter conservative rejection of under-specified wording, both of which cause the estimated precision to be a lower bound on \toolname{}'s true precision. 


As the invalid strata comprises only 4.3\% of the judgments, the overall precision remains high, despite the judge being conservative with the invalid strata.
Applying the stratified proportion estimator~\cite{scheaffer2012elementary}(Ch~5.6),
the estimated total number of valid behaviours ($V\!B_{est}$) surfaced by \toolname{} 
is:
\[
V\!B_{est} = n_{valid} \cdot \hat{p}_{valid} + n_{invalid} \cdot \hat{p}_{invalid} \hspace{0.6em}
\]
\vspace{-1.5em}
\[
V\!B_{est} = 19{,}829 \cdot 0.96 + 900 \cdot 0.28 = 19{,}288
\]
Precision is then: 
\[
\hat{P_{beh}} = \frac{VB_{est}}{AB} = \frac{19,\!288}{20,\!729} \approx 0.931
\]
To bound precision uncertainty, we apply a stratified variance estimator with finite population correction:
\[
\text{Var}(\hat{P_{beh}}) = \frac{1}{N^2} \sum_{h \in \{v,i\}} 
N_h^2 \cdot \frac{1 - f_h}{n_h} \cdot \hat{p}_h(1-\hat{p}_h)
\] where $f_h = n_h / N_h$ is the sampling fraction for stratum $h$. 
This yields standard error $\text{SE}(\hat{P_{beh}}) \approx 0.0266$.

\smallskip
\noindent\textbf{Implication:} This analysis shows that \toolname{} is able to reliably extract real behaviours from software methods and their documentation with high precision. 
The methodology also demonstrates how a carefully-developed and validated automated judge can be shown to have substantial inter-rater reliability compared to multiple human labellers, enabling analysis of whole datasets instead of small manual samples.

The judge's errors were not random. Failures in behaviour extraction concentrated in cases where behaviours were implied by context rather than explicitly stated, or where documentation was ambiguous. 
Both failure modes cause underestimation: implied behaviours that the judge rejects are real behaviours that do not enter the reference set; ambiguous behaviours flagged as invalid reduce the denominator of confirmed gaps. 
As new models are developed that could be used for the behaviour extractor or judge, more of these cases could be resolved correctly, increasing the estimated number of valid behaviours that can be checked against existing test cases.

\begin{rqbox}{RQ1: Precision of \toolname{}'s behaviour extraction.}
Across all evaluated projects, \toolname{} conservatively yields a behaviour detection precision of $93.1\%\pm2.7\%$. 
This high precision and low variance demonstrate that \toolname{} can consistently extract valid expected behaviours, establishing a reliable foundation for downstream test adequacy analysis.
\end{rqbox}

\subsection{RQ2 (Prevalence): Do developer-written test suites exhibit behavioural gaps in practice?}
\label{sec:rq2}


A test suite can execute every line of code and still leave expected behaviours un-validated by the test suite, as shown in Section~\ref{sec:example}. 
Using the 20,729 extracted behaviours from RQ1, we next investigated how prevalent these validation gaps were across the developer-written test suites in our ten project dataset.
Following the process in Figure~\ref{fig:tool}(c), \toolname{} identifies 5,083 untested behaviours (UTB) 
in the dataset.
In this research question we follow the same two-step process as RQ1: first, we establish human consensus labels to determine whether these gaps are valid. Second, we build a new LLM-based judge and compare its labelling performance against the prior human labels and use this to assess \toolname{}'s gap identification precision.
Third, we manually examined the gaps to learn which kinds of gaps exist in practice.

\smallskip
\subsubsection{How prevalent are behavioural gaps in developer-written test suites?} 
We first assess how precisely \toolname{} is able to assess behavioural gaps, and model their existence. 


\smallskip
\noindent\textbf{Gap assessment judge:} This judge is based on the same judge design as used in RQ1. 
For each untested behaviour, the judge is provided with the behaviour text, the method's documentation, implementation, and any test cases that directly or transitively invoke the method.
The judge then either labels the behaviour as \textit{already tested} by the existing test suite, or that the behaviour is \textit{untested} and represents a valid behavioural gap.
Across the 5,083 UTBs surfaced by \toolname{}, the judge classified 4,722 (92.9\%) behaviours as \textit{untested} and 361 (7.1\%) behaviours as \textit{already tested}.

\smallskip
\noindent\textbf{Validating gap judge against human labels:}
We validate the judge's output against human labellers before applying it at scale. 
We used the same disproportional stratified sampling approach as in RQ1. 
Two authors independently annotated 100 samples, drawing 50 from the judge's \textit{untested} stratum and 50 from the \textit{already tested} stratum, distributed across all 10 projects. 
The independent human labels achieved \emph{almost perfect} agreement (Cohen's $\kappa=0.83$).
Disagreements were resolved collaboratively to produce consensus labels, serving as the ground truth reference against which the judge's predictions are compared. 
We believe human consensus is the appropriate ground truth for these tasks as the goal is to interpret what a human developer would expect rather than what the documentation strictly states.
Comparing the gap assessment judge against these consensus labels yields $\kappa=0.66$, representing \emph{substantial agreement}. 
A post-hoc power analysis ($\alpha=0.05$, power = 0.90) indicates that our sample of 100 exceeds the minimum required threshold of 98. 



\smallskip
\noindent\textbf{Estimating UTB prevalence in developer-written suites:}
We applied the same stratified proportion estimator from RQ1 to obtain a lower bound on the true number of behavioural gaps. Based on the consensus labelling, in 
the \textit{untested behaviour} stratum, 38 out of 50 samples were confirmed as genuine gaps ($\hat{p}_{valid}$=0.76). In the \textit{already tested} stratum, 45 out of 50 samples were truly already tested, meaning 5 were missed gaps incorrectly classified as covered ($\hat{p}_{invalid}$=0.10).
Specifically, the judge tended to flag behaviours as untested when coverage was implicit rather than explicit, such as contracts enforced through method chaining or transitive test execution where a test exercises the behaviour without a direct assertion.
Using the stratified proportion estimator, the number of valid untested behaviours ($VUTB_{est}$) reported by \toolname{} is:
\[
VUTB_{est} =  4,\!722 \cdot 0.76 + 361 \cdot 0.10 = 3,\!625
\]

Precision is then: 
\[
\hat{P_{utb}} = \frac{VUTB_{est}}{UTB} = \frac{3,\!625}{5,\!083} \approx 0.713
\]


%
The standard error for this analysis is 0.056.
For the projects in the dataset we therefore estimate that $71.3\%\pm5.6\%$ of the untested behaviours identified by \toolname{} represent behaviours a developer would expect of a method, but are not tested by the existing test suite.
Applying this precision to the all behaviours (AB) identified in RQ1 yields:

\[
{UTB_{est}} = \frac{UTB}{AB} \cdot {P_{utb} = \frac{5,\!083}{20,\!729} \cdot {0.713} \approx 0.175} 
\]

This means that across the ten projects and their test suites in the dataset, 17.5\% of the detected expected behaviours represent valid behavioural gaps.
Behavioural gaps range from 6.6\%--29.3\% across projects, with jfreechart as the largest outlier. Removing jfreechart from the aggregate yields a gap of 15.1\%. 
Every project in the dataset exhibits gaps, including those with the strongest structural metrics (Table~\ref{tab:repositories}): commons-cli (98.2\% line, 93\% kill) still has a 6.6\% gap, and commons-csv (99.6\% line, 95\% kill) has 9.1\%. This is a project-level companion to RQ4's method-level result. 

\begin{table}[h]
\centering
\caption{Per-project lower bound estimates for the behavioural gaps across the dataset.
}
\label{tab:rq2_dev}
\begin{tabular}{lrrr}
\toprule
& \textbf{Detected} & \textbf{Lower Bound} & \textbf{Lower Bound} \\
\textbf{Project} & \textbf{UTB} & \textbf{UTB} & \textbf{Behavioural Gap} \\
\midrule
commons-cli         & 39    & 28    & 6.6\% \\
commons-codec       & 277   & 198   & 16.3\% \\
commons-compress    & 413   & 295   & 18.3\% \\
commons-jxpath      & 52    & 37    & 20.2\% \\
commons-collections & 690   & 492   & 17.5\% \\
commons-lang        & 1,387 & 989   & 12.8\% \\
commons-csv         & 46    & 33    & 9.1\% \\
gson                & 105   & 75    & 13.5\% \\
jfreechart          & 1,863 & 1,328 & 29.3\% \\
jsoup               & 211   & 150   & 12.3\% \\
\midrule
\textbf{Total}      & \textbf{5,083} & \textbf{3,625} & \textbf{17.5\%} \\
\bottomrule
\end{tabular}
\end{table}

\smallskip
\subsubsection{What kinds of behaviours do developers overlook in their test suites?} 
We further examined the UTBs to identify the kinds of behaviours developers frequently miss when constructing their existing test suites, despite these suites having high code coverage. 
To do this, we used an open coding grounded theory methodology. The authors iterated through a set of valid untested behaviours, identifying labels that summarized each behaviour.
These labels were merged iteratively until saturation was reached. 
The most common categories of untested behaviours in developer-written test suites, and an example behaviour of each, includes:

\begin{itemize}
    \item \textit{Exceptions:} Cases where methods are expected to consistently throw a specific exception under a given condition.
    For example, \textit{``Throws a \texttt{NullPointerException} when the input iterable is null.''}
    \item \textit{Side Effects:} Behaviour describing how a method alters internal state or data beyond what is directly returned. 
    For example, \textit{``Preserves existing entries in the original bag without applying the transformer.''}
    \item \textit{Handling null/empty inputs:} Explicit documentation of return values or behaviours for null inputs, empty inputs, or absent data.
    For example, \textit{``Returns \texttt{null} when the provided map does not contain the specified key.''}
    \item \textit{Construction \& configuration:} Rules describing how objects are initialized or configured. 
    For example, \textit{``Constructs a new, empty map with the specified maximum size and load factor.''} 
    \item \textit{Content of returned value:} Descriptions of the content or structure of the return value.
    For example, \textit{``Return the previous value of the passed element before it was updated.''}
    \item \textit{Algorithmic logic:} Descriptions of specific computational rules or constraints. For example, \textit{``Clamps the \texttt{endIndexExclusive} parameter to the array length if it exceeds the array length.''} 
\end{itemize}

\smallskip
\noindent\textbf{Implication:} The dataset consists of popular, well-maintained libraries with high structural coverage and kill scores.
This means the 17.5\% gap observed by \toolname{} is likely conservative and could be larger for projects that have made less extensive investments in their automated test suites.
Table~\ref{tab:rq2_dev} shows the per-project breakdown of the identified gaps. 
These results show that all ten projects exhibit untested behaviours, with a consistent direction (gaps range from 6.6\%--29.3\%). 

Notably, none of the categories of untested behaviours seem inherently untestable. Ensuring objects are constructed appropriately, handle errors correctly, have expected side effects, and return properly-updated values are all observable by standard automated testing approaches.
While exploring the categories, we found five 
gaps that pointed to inconsistencies and we reported them to their respective open-source projects, and all of them were fixed.


As in RQ1, the 17.5\% UTB rate represents a lower bound with respect to the capability of the models used by \toolname{} and the judge.
The 71.3\% precision already discounts cases where the mapper reported a gap for a behaviour the test suite in fact validated, so the 17.5\% estimate is not inflated by those false positives.
It cannot, however, account for untested behaviours the mapper wrongly treated as validated, nor for behaviours the extractor never surfaced; neither appears among the 5,083 candidates the judge examined. 
A more capable extractor would surface a higher number of valid behaviours, and a more capable mapper would mislabel fewer untested behaviours as validated, so stronger models move the corrected estimate upward rather than down.

\begin{rqbox}{RQ2: Prevalence of dev-written behavioural gaps.}
%
Across ten well-tested Java systems, at least 17.5\% of the detected behaviours are not validated by developer-written test suites. 
These gaps arise for a variety of reasons, none of which are because the behaviours are inherently untestable.
\end{rqbox}

\subsection{RQ3 (Generalization): To what extent are auto\-matically-generated test suites prone to behavioural gaps?}
\label{sec:rq3}

This research question seeks to investigate whether behavioural gaps are inherent to developer-written suites, or a systemic limitation of structure-guided testing practices.
To do this, we evaluate whether behavioural gaps persist in test suites created by automatic test generation approaches.
We chose two state‑of‑the‑art (SOTA) automated test generators: EvoSuite~\cite{10.1145/2025113.2025179} and ASTER~\cite{Pan2024ASTERNA}, to evaluate whether they can systematically close behavioural gaps.
ASTER was selected because it represents the SOTA for LLM-supported test generation, and is a shipping industrial product for automatically generating test suites.
EvoSuite was selected because it is the baseline test generation tool widely used in academia, and was also used by the ASTER team in their own comparative evaluation.
Evaluating whether automatically generated test suites would help contextualize whether behavioural gaps are just a byproduct of misplaced human effort, or a more pervasive testing challenge. 

\smallskip
\noindent\textbf{Methodology:} 
ASTER provides a robust replication package including both ASTER and EvoSuite generated tests for eight projects.
Unfortunately, ASTER is not publicly available so we restricted our analysis to the four projects in the ASTER dataset that intersected with the projects in our dataset.
To enable a fair comparison between the three tools, we only evaluated ASTER-\toolname{} on public documented methods ASTER was able to generate tests for. The same holds for the methods examined in the EvoSuite-\toolname{} comparison. 
This accounts for the different method counts in Table~\ref{tab:rq3}.

To ensure behavioural gaps were not just due to test suites having low structural metric values, we also computed the line coverage for the analyzed methods used for each test generation tool.
For ASTER, 81.2\% of methods achieve 91–100\% line coverage; for EvoSuite, 89.7\% achieve 91-100\% line coverage.

Since the four projects were already in our dataset, we already had the expected behaviours for all of the methods being analyzed, but \toolname{} had to be executed again with the test suites generated by ASTER and EvoSuite to identify the behavioural gaps in these suites.
For this analysis we only report the estimated number of behavioural gaps by multiplying the number of UTBs identified by \toolname{} by the 0.713 precision from RQ2.

\begin{table}[h]
\centering
\caption{UTB and method concentration across coverage buckets (aggregated across 
all projects). UTBs (\%) is the share of total UTBs falling in each bucket. 
Methods (\%) is the share of total methods in each bucket.}
\label{tab:rq4}
\small
\setlength{\tabcolsep}{3.5pt}
\begin{tabular}{l cc cc}
\toprule
& \multicolumn{2}{c}{\textbf{Line Coverage}} 
& \multicolumn{2}{c}{\textbf{Mutation Score}} \\
\cmidrule(lr){2-3} \cmidrule(lr){4-5}
\textbf{Bucket} 
& \textbf{UTBs (\%)} & \textbf{Methods (\%)} 
& \textbf{UTBs (\%)} & \textbf{Methods (\%)} \\
\midrule
$\leq$70\%  & 5.2  & 2.8  & 44.3 & 24.3 \\
71--80\%    & 4.1  & 2.4  & 3.0  & 3.1  \\
81--90\%    & 5.4  & 2.6  & 2.6  & 2.7  \\
91--100\%   & 85.4 & 92.3 & 50.1 & 69.8 \\
\bottomrule
\end{tabular}
\end{table}

\smallskip
\noindent\textbf{Estimating UTB prevalence in generated suites:}
The behavioural gaps left by both ASTER and EvoSuite are reported in Table~\ref{tab:rq3}.
In summary, 27.1\% of expected behaviours were not validated by the ASTER-generated suites and 20.6\% of the expected behaviours were not validated by the EvoSuite-generated suites.
These levels of behavioural gaps both exceed the developer-written suites for these four projects, as detailed in Table~\ref{tab:rq2_dev}.

This suggests that neither EvoSuite, a search-based test generator, nor ASTER, an LLM-based test generator are immune to missing expected behaviours in software systems.
One reason for this could be because neither ASTER nor EvoSuite use documentation to guide their test generation. 

We also examined Judot~\cite{denaro2025automatedtestgenerationprogram}, a documentation-aware test generator that uses JDoctor to extract pre- and postconditions from structured Javadoc tags. 
Across the four projects, it generated tests for only 149 methods, compared to 778 for ASTER and 1,787 for EvoSuite, since its parser is limited to formal tags and ignores narrative documentation. 
On the methods it did reach, the tests showed a 42.5\% behavioural gap, suggesting that even documentation-aware generation, when restricted to structured tags, captures only a fraction of the behaviours developers express in method-level documentation.

\begin{table}[!t]
\centering
\caption{Behavioural gap results for ASTER and EvoSuite across four projects. 
Beh. Gap\% is the conservative lower-bound estimate of confirmed behavioural gaps using the developer-written confirmation rate of 71.3\%.}
\label{tab:rq3}
\begin{tabular}{llrrr}
\toprule
\textbf{Project} & \textbf{Tool} & \textbf{Methods} & \textbf{UTBs} & \textbf{Beh. Gap\%} \\
\midrule
commons-cli      & ASTER    & 169   & 91  & 19.5\% \\
commons-codec    & ASTER    & 275 & 255 & 27.7\% \\
commons-compress & ASTER    & 200 & 165 & 31.7\% \\
commons-jxpath   & ASTER    & 134 & 108 & 28.4\% \\ \midrule
\textbf{Total}   & & \textbf{778} & \textbf{619} & \textbf{27.1\%} \\
\midrule
\midrule
commons-cli      & EvoSuite & 213  & 122 & 20.7\% \\
commons-codec    & EvoSuite & 470 & 318 & 20.8\% \\
commons-compress & EvoSuite & 812 & 440 & 22.3\% \\
commons-jxpath   & EvoSuite & 292 & 120 & 15.8\% \\ \midrule
\textbf{Total}   &  & \textbf{1,787} & \textbf{1,000} & \textbf{20.6\%} \\
\bottomrule
\end{tabular}
\vspace{-2em}
\end{table}
\smallskip
\noindent\textbf{Implication:} Both of the test generation approaches failed to have smaller behavioural gaps than the developer-written suites. 
This suggests that behavioural gaps do not arise due to developer inattention, but instead because they capture aspects of software validation that are orthogonal to the structural metrics test generation approaches traditionally optimize for.

\begin{rqbox}{RQ3: Prevalence of generated behavioural gaps.}
Behavioural gaps are not unique to developer‑written suites. Suites generated by ASTER and EvoSuite exhibit behavioural gaps of 27.1\% and 20.6\% respectively, despite achieving strong structural coverage on the methods they exercise.
\end{rqbox}

\subsection{RQ4 (Implication): Do behavioural gaps alias traditional structural metrics?}
\label{sec:rq4}

Line coverage and mutation kill score are the two most common proxies for test suite adequacy~\cite{PAPADAKIS2019275}. Line coverage records whether source lines are executed; mutation kill score additionally requires that assertions detect injected faults. If either metric already captured behavioural gaps, a method with a strong structural score would contain few untested behaviours, and identifying behavioural gaps would add nothing beyond what these metrics already report. 
We test this for each metric in turn by asking whether a method's structural score predicts its behavioural coverage, and whether untested behaviours disappear once that score is high. We report Spearman's $\rho$ (behavioural coverage is not normally distributed)~\cite{spearman1961proof} and the variance explained by OLS $R^2$, and we measure how strongly untested behaviours concentrate in low-scoring methods with Cohen's $h$, comparing the share of UTBs in the top score bucket against the share of methods in that bucket. Table~\ref{tab:rq4} reports the bucket distribution for developer-written suites in the dataset.

\smallskip
\noindent\textbf{Line coverage:} 
Line coverage confers almost no information about behavioural gaps. 
Its correlation with behavioural coverage is negligible ($\rho=0.112, R^2=0.007$).
The distribution of UTBs across coverage buckets nearly matches the distribution of methods: 85.4\% of UTBs fall in the 91–100\% bucket along with 92.3\% of all methods. 
The effect size of the difference is considered \textit{small} (Cohen's $h=0.22$). 
High coverage does not eliminate gaps; among methods with 100\% line coverage, 38.2\% still contain at least one untested behaviour. 
A method can execute every line under test and still leave documented behaviour untested, as demonstrated by the \texttt{emptyArray()} example in Section~\ref{sec:example}.

\smallskip
\noindent\textbf{Mutation kill score:} 
Killing a mutant confers more information because killing a mutant requires assertions to detect behaviours rather than only executing code. 
Kill score correlates more strongly with behavioural coverage ($\rho=0.377$).
Untested behaviours are more prevalent in methods with low kill score (44.3\% of UTBs fall in the $\leq70\%$ kill score bucket containing only 24.3\% of methods); the effect size of the relationship is considered \textit{medium} (Cohen's $h=0.41$).
Although a stronger relationship exists than for coverage, kill score explains only 16\% of the variance in behavioural coverage ($R^2=0.162$), leaving most of the difference unaccounted for.

High kill scores also do not eliminate behavioural gaps as 29.6\% of methods with a perfect mutation kill score contain at least one untested behaviour. 
A strong kill score lowers the odds of a behavioural gap without ruling one out. 
For both metrics, a chi-square test of homogeneity confirms that the distribution of UTBs across buckets differs significantly from the distribution of methods ($p<0.001$).

\smallskip
\noindent\textbf{Implication:} 
The same observations above hold for the automatically generated suites. 
Among methods with a perfect structural score, using ASTER 43\% (coverage) / 38\% (kill score) still contain untested behaviours; using EvoSuite, 32\% (coverage) / 34\% (kill score) also contain untested behaviours, so the gap does not close even with generated test suites.

Neither structural metric certifies that a method's documented behaviour is validated. Line coverage is effectively uninformative about behavioural gaps, and mutation kill score is only a weak signal that still misses gaps in nearly a third of methods with a perfect kill score.
Behavioural coverage is therefore a complementary dimension of test adequacy rather than a restatement of either metric. We note that this analysis bounds the relationship from one side only: the 17.5\% figure is a lower bound on confirmed gaps, because \toolname{} is not guaranteed to surface every untested behaviour. 
Whether additional gaps go undetected is left to future evaluation.

\begin{rqbox}{RQ4: Do UTBs alias existing structural metrics?}
Line coverage is effectively uninformative about behavioural gaps, and mutation kill score is only a weak signal: for developer-written tests, 38.2\% of methods with perfect line coverage and 29.6\% with a perfect kill score still contain untested behaviours. This suggests that behavioural coverage is a complementary dimension of test adequacy. 
\end{rqbox}

\section{Conclusion}
\label{sec:summary}

Software testing research has long relied on structural metrics as proxies for test suite quality, partly because reliable approaches do not exist for measuring whether a test suite validates a method's documented behaviour.
In this paper, we introduce the \toolname{} proof-of-concept that extracts method-level expected behaviours from documentation and source code, maps them to test cases, and identifies which documented behaviours go untested. 
Across 8,922 methods in ten well-tested Java libraries, \toolname{} surfaces 20,729 behaviours at $93.1\%\pm2.7\%$ precision. At least 17.5\% of those detected behaviours are untested in developer-written suites. Test suites generated by ASTER / EvoSuite leave 27.1\% / 20.6\% of detected behaviours untested, both exceeding developer-written behavioural gaps on the same projects. 
Qualitative inspection of the missed behaviours (exception handling, side effects, null input handling) confirms that none are inherently untestable.
Behavioural gaps provide useful independent insight from structural metrics: Among methods with perfect line coverage in our dataset, 38.2\% still contain at least one behavioural gap; among methods with perfect mutation kill scores, 29.6\% do. Spearman correlations of 0.112 (line) and 0.377 (mutation) confirm that neither metric reliably predicts behavioural gaps. 
Ultimately, 
coverage reports and kill scores tell developers how thoroughly the implementation has been exercised, but behavioural gap analysis can provide novel test adequacy insight into whether code was validated consistently with its expected behaviour.


\balance
\bibliographystyle{IEEEtran}
\bibliography{references}

@inproceedings{10.1145/2025113.2025179,
author = {Fraser, Gordon and Arcuri, Andrea},
title = {{EvoSuite: A}utomatic test suite generation for object-oriented software},
year = {2011},
doi = {10.1145/2025113.2025179},
booktitle = {Proceedings of the European Conference on Foundations of Software Engineering (ESEC)},
pages = {416--419},
numpages = {4},
}

@misc{hossain2025doc2oracllinvestigatingimpactdocumentation,
      title={{Doc2OracLL: I}nvestigating the Impact of Documentation on {LLM}-based Test Oracle Generation}, 
      author={Soneya Binta Hossain and Raygan Taylor and Matthew Dwyer},
      year={2025},
      eprint={2412.09360},
      archivePrefix={arXiv},
      primaryClass={cs.SE},
      url={https://arxiv.org/abs/2412.09360}, 
}

@inproceedings{10.1145/3213846.3213872,
author = {Blasi, Arianna and Goffi, Alberto and Kuznetsov, Konstantin and Gorla, Alessandra and Ernst, Michael D. and Pezz\`{e}, Mauro and Castellanos, Sergio Delgado},
title = {Translating code comments to procedure specifications},
year = {2018},
doi = {10.1145/3213846.3213872},
booktitle = {Proceedings of the International Symposium on Software Testing and Analysis (ISSTA)},
pages = {242--253},
numpages = {12},
}

@misc{denaro2025automatedtestgenerationprogram,
      title={Automated Test Generation from Program Documentation Encoded in Code Comments}, 
      author={Giovanni Denaro and Luca Guglielmo},
      year={2025},
      eprint={2504.21161},
      archivePrefix={arXiv},
      primaryClass={cs.SE},
      url={https://arxiv.org/abs/2504.21161}, 
}

@inproceedings{10.1145/2931037.2931061,
author = {Goffi, Alberto and Gorla, Alessandra and Ernst, Michael D. and Pezz\`{e}, Mauro},
title = {Automatic generation of oracles for exceptional behaviors},
year = {2016},
doi = {10.1145/2931037.2931061},
booktitle = {Proceedings of the International Symposium on Software Testing and Analysis (ISSTA)},
pages = {213--224},
numpages = {12},
}

@inproceedings{blasi2022call,
  title={Call me maybe: {Using NLP} to automatically generate unit test cases respecting temporal constraints},
  author={Blasi, Arianna and Gorla, Alessandra and Ernst, Michael D and Pezz{\`e}, Mauro},
  booktitle={Proceedings of the International Conference on Automated Software Engineering (ASE)},
  pages={1--11},
  year={2022}
}

@article{kurian2023automatically,
  title={Automatically generating test cases for safety-critical software via symbolic execution},
  author={Kurian, Elson and Briola, Daniela and Braione, Pietro and Denaro, Giovanni},
  journal={Journal of Systems and Software},
  volume={199},
  pages={111629},
  year={2023},
  publisher={Elsevier}
}

@inproceedings{pacheco2007feedback,
  title={Feedback-directed random test generation},
  author={Pacheco, Carlos and Lahiri, Shuvendu K and Ernst, Michael D and Ball, Thomas},
  booktitle={Proceedings of the International Conference on Software Engineering (ICSE)},
  pages={75--84},
  year={2007},
}

@misc{qwen3technicalreport,
      title={Qwen3 Technical Report}, 
      author={Qwen Team},
      year={2025},
      eprint={2505.09388},
      archivePrefix={arXiv},
      primaryClass={cs.CL},
      url={https://arxiv.org/abs/2505.09388}, 
}

@misc{zhang2025frameworkcreatingnonregressivetest,
      title={A Framework for Creating Non-Regressive Test Cases via Branch Consistency Analysis Driven by Descriptions}, 
      author={Yuxiang Zhang and Pengyu Xue and Zhen Yang and Xiaoxue Ren and Xiang Li and Linhao Wu and Jiancheng Zhao and Xingda Yu},
      year={2025},
      eprint={2506.07486},
      archivePrefix={arXiv},
      primaryClass={cs.SE},
      url={https://arxiv.org/abs/2506.07486}, 
}

@incollection{papadakis2019mutation,
  title={Mutation testing advances: an analysis and survey},
  author={Papadakis, Mike and Kintis, Marinos and Zhang, Jie and Jia, Yue and Le Traon, Yves and Harman, Mark},
  booktitle={Advances in computers},
  volume={112},
  pages={275--378},
  year={2019},
  publisher={Elsevier}
}

@article{design-tradeoffs,
author = {Robillard, Martin P. and Arya, Deeksha M. and Ernst, Neil A. and Guo, Jin L. C. and Lamothe, Maxime and Nassif, Mathieu and Novielli, Nicole and Serebrenik, Alexander and Steinmacher, Igor and Stol, Klaas-Jan},
title = {Communicating Study Design Trade-offs in Software Engineering},
year = {2024},
issue_date = {June 2024},
volume = {33},
number = {5},
doi = {10.1145/3649598},
journal = {ACM Transactions on Software Engineering Methodology (TOSEM)},
month = jun,
articleno = {112},
numpages = {10}
}

@inproceedings{10.1145/3611643.3616327,
author = {Davis, Matthew C. and Choi, Sangheon and Estep, Sam and Myers, Brad A. and Sunshine, Joshua},
title = {NaNofuzz: A Usable Tool for Automatic Test Generation},
year = {2023},
isbn = {9798400703270},
publisher = {Association for Computing Machinery},
address = {New York, NY, USA},
url = {https://doi.org/10.1145/3611643.3616327},
doi = {10.1145/3611643.3616327},
pages = {1114–1126},
numpages = {13},
location = {San Francisco, CA, USA},
series = {ESEC/FSE 2023}
}

@inproceedings{maton2025tests,
  title={Where tests fall short: empirically analyzing oracle gaps in covered code},
  author={Maton, Megan and Kapfhammer, Gregory M and McMinn, Phil},
  booktitle={Proceedings of the International Symposium on Empirical Software Engineering and Measurement (ESEM 2025)},
  year={2025},
  organization={Institute of Electrical and Electronics Engineers (IEEE)}
}

@misc{xu2025docprismlocalcategorizationexternal,
      title={DocPrism: Local Categorization and External Filtering to Identify Relevant Code-Documentation Inconsistencies}, 
      author={Xiaomeng Xu and Zahin Wahab and Reid Holmes and Caroline Lemieux},
      year={2025},
      eprint={2511.00215},
      archivePrefix={arXiv},
      primaryClass={cs.SE},
      url={https://arxiv.org/abs/2511.00215}, 
}

@article{schuler2013checked,
author = {Schuler, David and Zeller, Andreas},
title = {Checked coverage: an indicator for oracle quality},
journal = {Software Testing, Verification and Reliability},
volume = {23},
number = {7},
pages = {531-551},
doi = {https://doi.org/10.1002/stvr.1497},
url = {https://onlinelibrary.wiley.com/doi/abs/10.1002/stvr.1497},
eprint = {https://onlinelibrary.wiley.com/doi/pdf/10.1002/stvr.1497},
year = {2013}
}

@article{vera2019comprehensive,
author = {Vera Pérez, Oscar and Danglot, Benjamin and Monperrus, Martin and Baudry, Benoit},
year = {2019},
month = {06},
pages = {},
title = {A Comprehensive Study of Pseudo-tested Methods},
volume = {24},
journal = {Empirical Software Engineering},
doi = {10.1007/s10664-018-9653-2}
}

@article{mchugh2012interrater,
  title={Interrater reliability: the kappa statistic},
  author={McHugh, Mary L},
  journal={Biochemia medica},
  volume={22},
  number={3},
  pages={276--282},
  year={2012},
  publisher={Hrvatsko dru{\v{s}}tvo za medicinsku biokemiju i laboratorijsku medicinu}
}

@inproceedings{10.1145/2635868.2635929,
author = {Just, Ren\'{e} and Jalali, Darioush and Inozemtseva, Laura and Ernst, Michael D. and Holmes, Reid and Fraser, Gordon},
title = {Are mutants a valid substitute for real faults in software testing?},
year = {2014},
doi = {10.1145/2635868.2635929},
booktitle = {Proceedings of the International Symposium on Foundations of Software Engineering (ESEC/FSE)},
pages = {654--665},
numpages = {12},
}

@misc{pitest2026,
  title={{PIT: Mutation Testing for Java}},
  author={{PIT Contributors}},
  year={2026},
  howpublished={\url{https://pitest.org/}},
  note={Accessed: 2025-12-22}
}

@misc{jacoco2026,
  title={{JaCoCo Java Code Coverage Library}},
  author={{EclEmma and JaCoCo Contributors}},
  year={2026},
  howpublished={\url{https://www.jacoco.org/jacoco/trunk/doc/maven.html}},
  note={Accessed: 2025-12-25}
}

@article{Wong_2023,
   title={Natural Language Generation and Understanding of Big Code for AI-Assisted Programming: {A} Review},
   volume={25},
   DOI={10.3390/e25060888},
   number={6},
   journal={Entropy},
   author={Wong, Man-Fai and Guo, Shangxin and Hang, Ching-Nam and Ho, Siu-Wai and Tan, Chee-Wei},
   year={2023},
   month=jun, pages={888} }

@inproceedings{chen2024chatunitest,
  title={Chatunitest: A framework for llm-based test generation},
  author={Chen, Yinghao and Hu, Zehao and Zhi, Chen and Han, Junxiao and Deng, Shuiguang and Yin, Jianwei},
  booktitle={Companion Proceedings of the 32nd ACM International Conference on the Foundations of Software Engineering},
  pages={572--576},
  year={2024}
}

@inproceedings{10.1109/ICST.2012.110,
author = {Fraser, Gordon and Walkinshaw, Neil},
title = {Behaviourally Adequate Software Testing},
year = {2012},
doi = {10.1109/ICST.2012.110},
booktitle = {Proceedings International Conference on Software Testing, Verification and Validation (ICST)},
pages = {300--309},
numpages = {10},
}

@article{Pan2024ASTERNA,
  title={ASTER: Natural and Multi-Language Unit Test Generation with LLMs},
  author={Rangeet Pan and Myeongsoo Kim and Rahul Krishna and Raju Pavuluri and Saurabh Sinha},
  journal={Proceedings of the International Conference on Software Engineering: Software Engineering in Practice (ICSE-SEIP)},
  year={2024},
  pages={413--424}
}

@article{Endres2023CanLL,
  title={Can Large Language Models Transform Natural Language Intent into Formal Method Postconditions?},
  author={Madeline Endres and Sarah Fakhoury and Saikat Chakraborty and Shuvendu K. Lahiri},
  journal={Proceedings of the ACM on Software Engineering},
  year={2023},
  volume={1},
  pages={1889--1912}
}

@inproceedings{10.1145/2610384.2628055,
author = {Just, Ren\'{e} and Jalali, Darioush and Ernst, Michael D.},
title = {{Defects4J: A} database of existing faults to enable controlled testing studies for Java programs},
year = {2014},
doi = {10.1145/2610384.2628055},
booktitle = {Proceedings of the International Symposium on Software Testing and Analysis (ISSTA)},
pages = {437--440},
numpages = {4},
}

@incollection{PAPADAKIS2019275,
title = {Mutation Testing Advances: {A}n Analysis and Survey},
editor = {Atif M. Memon},
series = {Advances in Computers},
publisher = {Elsevier},
volume = {112},
pages = {275--378},
year = {2019},
doi = {10.1016/bs.adcom.2018.03.015},
author = {Mike Papadakis and Marinos Kintis and Jie Zhang and Yue Jia and Yves Le Traon and Mark Harman}
}

@inproceedings{10.1145/1062455.1062530,
author = {Andrews, J. H. and Briand, L. C. and Labiche, Y.},
title = {Is mutation an appropriate tool for testing experiments?},
year = {2005},
doi = {10.1145/1062455.1062530},
booktitle = {Proceedings of the International Conference on Software Engineering (ICSE)},
pages = {402--411},
numpages = {10},
}

@book{spearman1961proof,
  title={The proof and measurement of association between two things.},
  author={Spearman, Charles},
  year={1961},
  publisher={Appleton-Century-Crofts}
}

@misc{liu2026ministral,
      title={Ministral 3}, 
      author={Alexander H. Liu and Kartik Khandelwal and Sandeep Subramanian and Victor Jouault and Abhinav Rastogi and others},
      year={2026},
      eprint={2601.08584},
      archivePrefix={arXiv},
      primaryClass={cs.CL},
      url={https://arxiv.org/abs/2601.08584}, 
}

@inproceedings{ahmed2025can,
  title={Can LLMs replace manual annotation of software engineering artifacts?},
  author={Ahmed, Toufique and Devanbu, Premkumar and Treude, Christoph and Pradel, Michael},
  booktitle={Proceedings of the International Conference on Mining Software Repositories (MSR)},
  pages={526--538},
  year={2025},
}

@article{10.1145/3797276,
author = {He, Junda and Shi, Jieke and Zhuo, Terry Yue and Treude, Christoph and Sun, Jiamou and Xing, Zhenchang and Du, Xiaoning and Lo, David},
title = {LLM-as-a-Judge for Software Engineering: Literature Review, Vision, and the Road Ahead},
year = {2026},
doi = {10.1145/3797276},
journal = {Transactions on Software Engineering and Methodology (TOSEM)},
month = Feb,
}

@article{zhou2025llm,
  title={An llm-as-judge metric for bridging the gap with human evaluation in se tasks},
  author={Zhou, Xin and Kim, Kisub and Zhang, Ting and Weyssow, Martin and Gomes, Luis F and Yang, Guang and Liu, Kui and Xia, Xin and Lo, David},
  journal={arXiv preprint arXiv:2505.20854},
  year={2025}
}

@article{doi:10.1177/02537176261422290,
  title = {A Tutorial on Sample Size Calculation for Inter-rater and Intra-rater Agreement Studies},
  DOI = {10.1177/02537176261422290},
  journal = {Indian Journal of Psychological Medicine},
  author = {Madadizadeh,  Farzan and Bahariniya,  Sajjad},
  year = {2026},
  month = Feb 
}

@Article{rs11020185,
AUTHOR = {A. Ramezan, Christopher and A. Warner, Timothy and E. Maxwell, Aaron},
TITLE = {Evaluation of Sampling and Cross-Validation Tuning Strategies for Regional-Scale Machine Learning Classification},
JOURNAL = {Remote Sensing},
VOLUME = {11},
YEAR = {2019},
NUMBER = {2},
ARTICLE-NUMBER = {185},
URL = {https://www.mdpi.com/2072-4292/11/2/185},
ISSN = {2072-4292},
DOI = {10.3390/rs11020185}
}

@inbook{scheaffer2012elementary,
  author    = {Scheaffer, Richard L. and Mendenhall, III, William and Ott, R. Lyman and Gerow, Kenneth G.},
  title     = {Elementary Survey Sampling},
  chapter   = {5.6},
  edition   = {7th},
  publisher = {Brooks/Cole},
  year      = {2012},
}

@inproceedings{10.1145/2568225.2568278,
author = {Gopinath, Rahul and Jensen, Carlos and Groce, Alex},
title = {Code coverage for suite evaluation by developers},
year = {2014},
doi = {10.1145/2568225.2568278},
booktitle = {Proceedings of the International Conference on Software Engineering (ICSE)},
pages = {72--82},
numpages = {11},
}

@inproceedings{10.1145/2568225.2568271,
author = {Inozemtseva, Laura and Holmes, Reid},
title = {Coverage is not strongly correlated with test suite effectiveness},
year = {2014},
doi = {10.1145/2568225.2568271},
booktitle = {Proceedings International Conference on Software Engineering (ICSE)},
pages = {435--445},
numpages = {11},
}

@article{Howden1978FunctionalPT,
  title={Functional Program Testing},
  author={William E. Howden},
  journal={Transactions on Software Engineering (TSE)},
  year={1978},
  volume={SE-6},
  pages={162--169}
}

@inproceedings{lee2025metamon,
  title={Metamon: {F}inding inconsistencies between program documentation and behavior using metamorphic LLM queries},
  author={Lee, Hyeonseok and An, Gabin and Yoo, Shin},
  booktitle={International Workshop on Large Language Models for Code (LLM4Code)},
  pages={120--127},
  year={2025},
}

@article{goodenough1975toward,
  title = {Toward a theory of test data selection},
  volume = {10},
  ISSN = {1558-1160},
  DOI = {10.1145/390016.808473},
  number = {6},
  journal = {ACM SIGPLAN Notices},
  author = {Goodenough,  John B. and Gerhart,  Susan L.},
  year = {1975},
  month = Apr,
  pages = {493--510},
  notes = {They note that structural testing is fundamentally incapable of catching "requirements errors" (where the program doesn't do what it's supposed to do) and "errors of omission" (where a specified feature is completely missing). They argue that to know if a program will "successfully fulfill its specifications," test selection criteria must be derived directly from the specification, not the code structure.}
}

@article{hamlet1994foundations,
  title = {Foundations of software testing: dependability theory},
  volume = {19},
  DOI = {10.1145/195274.195400},
  number = {5},
  journal = {SIGSOFT Software Engineering Notes (SEN)},
  author = {Hamlet,  Dick},
  year = {1994},
  month = Dec,
  pages = {128--139},
  notes = {Hamlet explicitly differentiates between testing the program (what the developer built) and testing the specification (what the system is required to do). He argues that code coverage acts as a metric of how thoroughly we have examined our own implementation choices, but says nothing about whether we have built the right thing.}
}

@inproceedings{zhang2015assertions,
  author    = {Zhang, Yandong and Mesbah, Ali},
  title     = {Assertions are strongly correlated with test suite effectiveness},
  booktitle = {Proceedings of the Joint Meeting on Foundations of Software Engineering (ESEC/FSE)},
  pages     = {214--224},
  year      = {2015},
  doi       = {10.1145/2786805.2786858}
}

@inproceedings{petrovic2021does,
  author    = {Petrovi{\'c}, Goran and Ivankovi{\'c}, Marko and Fraser, Gordon and Just, Ren{\'e}},
  title     = {Does mutation testing improve testing practices?},
  booktitle = {Proceedings of the International Conference on Software Engineering (ICSE)},
  pages     = {910--921},
  year      = {2021},
  doi       = {10.1109/icse43902.2021.00087}
}

@article{gay2015risks,
  author  = {Gay, Gregory and Staats, Matt and Whalen, Michael and Heimdahl, Mats P. E.},
  title   = {The risks of coverage-directed test case generation},
  journal = {Transactions on Software Engineering (TSE)},
  volume  = {41},
  number  = {8},
  pages   = {803--819},
  year    = {2015},
  doi     = {10.1109/tse.2015.2421011}
}

@article{zhu1997software,
  author     = {Zhu, Hong and Hall, Patrick A. V. and May, John H. R.},
  title      = {Software unit test coverage and adequacy},
  journal    = {ACM Computing Surveys (CSUR)},
  volume     = {29},
  number     = {4},
  pages      = {366--427},
  year       = {1997},
  doi        = {10.1145/267580.267590}
}

@article{miller1963systematic,
  author  = {Miller, Joan C. and Maloney, Clifford J.},
  title   = {Systematic mistake analysis of digital computer programs},
  journal = {Communications of the ACM (CACM)},
  volume  = {6},
  number  = {2},
  pages   = {58--63},
  year    = {1963},
  doi     = {10.1145/366246.366248},
  notes = {This is the original coverage paper (path coverage more than statement coverage), but does not look as much at the code as one would like for a direct citation}
}

@article{howden1976reliability,
  title = {Reliability of the Path Analysis Testing Strategy},
  volume = {SE-2},
  DOI = {10.1109/tse.1976.233816},
  number = {3},
  journal = {Transactions on Software Engineering (TSE)},
  author = {Howden,  W.E.},
  year = {1976},
  month = Sept,
  pages = {208--215},
  notes= {this is a very early path coverage paper that is looking directly at the source code to determine paths}
}

@inproceedings{Budd1978Design,
  title     = {The Design of a Prototype Mutation System for Program Testing},
  author    = {Budd, Timothy A. and Lipton, Richard J. and Sayward, Frederick G. and DeMillo, Richard A.},
  booktitle = {Proceedings of the National Computer Conference},
  pages     = {623--627},
  year      = {1978},
 notes = {This is the paper that introduced the idea of mutation testing}
}

@inproceedings{richardson89,
  author    = {Richardson, Debra J. and O'Malley, Owen and Tittle, Cindy},
  title     = {Approaches to specification-based testing},
  booktitle = {Proceedings of the Symposium on Testing, Analysis, and Verification (TAV3)},
  pages     = {86--96},
  year      = {1989},
  doi       = {10.1145/75308.75319}
}

@incollection{juristo2003functional,
  title = {Functional Testing,  Structural Testing and Code Reading: {W}hat Fault Type Do They Each Detect?},
  booktitle = {Empirical Methods and Studies in Software Engineering},
  publisher = {Springer Berlin Heidelberg},
  author = {Juristo,  Natalia and Vegas,  Sira},
  year = {2003},
  pages = {208--232},
  DOI = {10.1007/978-3-540-45143-3_12}
}

@inproceedings{lawrance2005how,
  title = {How Well Do Professional Developers Test with Code Coverage Visualizations? An Empirical Study},
  DOI = {10.1109/vlhcc.2005.44},
  booktitle = {Symposium on Visual Languages and Human-Centric Computing (VL/HCC)},
  author = {Lawrance,  J. and Clarke,  S. and Burnett,  M. and Rothermel,  G.},
  pages = {53--60}
}

@article{zheng2023judging,
  title={Judging {LLM}-as-a-judge with {MT}-bench and chatbot arena},
  author={Zheng, Lianmin and Chiang, Wei-Lin and Sheng, Ying and Zhuang, Siyuan and Wu, Zhanghao and Zhuang, Yonghao and Lin, Zi and Li, Zhuohan and Li, Dacheng and Xing, Eric and others},
  journal={Advances in neural information processing systems},
  volume={36},
  pages={46595--46623},
  year={2023}
}

@book{myers1979art,
  author    = {Myers, Glenford J.},
  title     = {The Art of Software Testing},
  year      = {1979},
  publisher = {John Wiley \& Sons}
}

\end{document}